\documentclass[useAMS,usenatbib]{mn2e}

\usepackage{times}
\usepackage{graphics,epsfig}
\usepackage{graphicx}
\usepackage{amsmath}
\usepackage{amssymb}
\usepackage{bm}
\usepackage{dcolumn}
\usepackage{epsfig}
\usepackage{subfig}
\usepackage{tabularx}
\usepackage[usenames,dvipsnames,svgnames,table]{xcolor}

\title[Mass-metallicity relation of GRB hosts]{On the mass-metallicity relation, velocity dispersion and
gravitational well depth of GRB host galaxies}
\author[Arabsalmani et al.]
{Maryam Arabsalmani$^{1,2}$\thanks{E-mail: marabsal@eso.org}, Palle M\o ller$^{	1}$, Johan P. U. Fynbo$^{2}$,  Lise Christensen$^{2}$,  
\newauthor
Wolfram Freudling$^{1}$,
Sandra Savaglio$^{1,3,4}$,
Tayyaba Zafar$^{1}$
       \\
        $^{1}$European Southern Observatory, Karl-Schwarzschild-Strasse 2, 85748 Garching bei M\"{u}nchen, Germany\\
	$^{2}$Dark Cosmology Centre, Niels Bohr Institute, University of Copenhagen, Juliane Maries Vej 30, DK-2100 Copenhagen \O, Denmark\\
	$^{3}$Max-Planck-Institut f\"{u}r extraterrestrische Physik, Giessenbachstrasse 1, 85748 Garching bei M\"{u}nchen, Germany\\
	$^{4}$Physics Department, University of Calabria, via P. Bucci, I-87036 Arcavacata di Rende, Italy
}

\begin{document}
\date{}

\pagerange{\pageref{firstpage}--\pageref{lastpage}} \pubyear{}

\maketitle

\label{firstpage}

\begin{abstract}
We analyze a sample of 16 absorption systems intrinsic to long
duration GRB host galaxies at $z\gtrsim 2$
for which the metallicities are known. We compare the relation
between the metallicity and cold gas velocity width  for this sample 
to that of the QSO-DLAs,
and find complete agreement. 
We then compare the redshift evolution of the mass-metallicity relation of our sample to
that of QSO-DLAs and find that also GRB hosts favour a late onset of this
evolution, around a redshift of $\approx 2.6$. We compute predicted
stellar masses for the GRB host galaxies using the prescription
determined from QSO-DLA samples and compare the measured stellar
masses for the four hosts where stellar masses have been determined
from SED fits.  We find excellent agreement and conclude that, on
basis of all available data and tests, long duration GRB-DLA hosts and
intervening QSO-DLAs are consistent with being drawn from the same
underlying population.

GRB host galaxies and QSO-DLAs  are found to have different impact parameter
distributions and we briefly discuss how this may affect statistical
samples. The impact parameter distribution has two effects. First any
metallicity gradient will shift the measured metallicity away from the
metallicity in the centre of the galaxy, second the path of the
sightline through different parts of the potential well of the dark matter halo
will cause different velocity fields to be sampled. We report evidence
suggesting that this second effect may have been detected.
\end{abstract}


\section{Introduction}
\label{int}
Stellar mass and metallicity are two of the most fundamental physical
properties of galaxies. The metal enrichment of the inter stellar
medium (ISM) of a galaxy is a consequence of supernova explosions and
stellar winds and is therefore related to the star formation history
(SFH) of the galaxy. Also, the amount of mass in stars is a function
of galaxy SFH. Therefore, understanding the evolution of the two
properties and the relation between them is fundamental to understand
the formation and evolution of galaxies.  Observations have shown that
for local and low redshift galaxies a tight relation exists between
the galaxy mass and its metallicity \citep[see e.g.][]{Dekel03, Tremonti04}. 
The evolution of the mass-metallicity (MZ) relation has been
studied using emission lines from HII regions in galaxies out to
$z\gtrsim 3$ and reveal for a galaxy of a given stellar mass a trend
of a decreasing metallicity with increasing redshift 
\citep{Savaglio05, Erb06, Maiolino08, Troncos14}. 
Whether the numerous galaxies at the faint end of the
luminosity function follow extrapolations of the MZ relation cannot
easily be addressed using emission selected samples, and most studies
have focused on composite spectra or a few individually selected
massive galaxies \citep{Erb06, Henry13, Cullen14}. 
Gravitational lensing which magnifies the flux of
background sources is a powerful tool to probe the faint end of the
galaxy luminosity function. Studies of low-mass gravitationally lensed galaxies hint at a weaker
evolution of the MZ relation and an increasing scatter out to $z=2$
\citep{Richard11, Wuyts12, Christensen12, Belli13}. 

\begin{table*}
\begin{minipage}{147mm}
\caption{Sample of $20$ GRB host DLA systems with their metallicities and selected low ion lines used for 
velocity width measurements. 
Metallicity measurements are based on element X.  
References for redshifts, $HI$ column densities, metallicity measurements, 
and published low-ionization profiles are quoted in the table footnote.} 
\label{Tab1}
\begin{tabular}{@{}llccllll}
\hline
 &  &  &  &  &  &  &  \\
GRB & redshift & $\log{[N_{HI}/(cm^{-2})]}$ & $[X/H]$ & $X$ & selected line$^a$ & Instrument & Ref.$^b$ \\
 &  &  &  &  &  &  &  \\
\hline
000926	           &   $2.0379 $   &  $21.3 \pm0.2 $   &   $-0.13\pm0.21$ &   Zn      &  SiII 1808	               &Keck/ESI           & (1),(2)      \\   
030323	           &   $3.3718 $   &  $21.90\pm0.07$   &   $-1.26\pm0.20$ &   S       &  SII 1253	               &UT4/FORS2          & (3)      \\
050401	           &   $2.8992 $   &  $22.60\pm0.30$   &   $-1.0 \pm 0.4$ &   Zn      &  SiII 1808	 	       &VLT/FORS2          & (4)      \\
050505	           &   $4.2748 $   &  $22.05\pm0.10$   &   $-1.2 \pm... $ &   S       &  SiII 1527		       &Keck/LRIS          &  (5)     \\
050730	           &   $3.9686$    &  $22.10\pm0.10$   &   $-2.18\pm0.11$ &   S       &  NiII 1370	               &VLT/UVES           & (6)     \\
050820A	           &   $2.6147 $   &  $21.05\pm0.10$   &   $-0.39\pm0.10$ &   Zn      &  NiII 1741$^c$		       &VLT/UVES           &  (6),(7)    \\
050922C	           &   $2.1992 $   &  $21.4 \pm0.3 $   &   $-1.82\pm0.11$ &   Si      &  FeII 1608	               &VLT/UVES           &  (6),(8)   \\
060206	           &   $4.0480 $   &  $20.85\pm0.10$   &   $-0.84\pm0.10$ &   S       &  SII 1253	               &WHT/ISIS           &  (9)     \\
060510B            &   $4.941  $   &  $21.3 \pm0.1 $   &   $-0.85\pm0.15$ &   S       &  NiII 1317	               &Gemini/GMOS        &  (10)     \\
070802	           &   $2.4549 $   &  $21.50\pm0.20$   &   $-0.50\pm0.68$ &   Zn      &  SiII 1808	               &VLT/FORS2          &  (11) \\
071031	           &   $2.6922 $   &  $22.15\pm0.05$   &   $-1.73\pm0.05$ &   Zn      &  NiII 1317$^d$		       &VLT/UVES           &  (6),(7) \\
080210	           &   $2.641  $   &  $21.90\pm0.10$   &   $-1.21\pm0.16$ &   Si      &  SiII 1808	               &VLT/FORS2          &  (12)     \\
081008	           &   $1.9683 $   &  $21.11\pm0.10$   &   $-0.52\pm0.11$ &   Zn      &  CrII 2056	               &VLT/UVES           &  (13)     \\
090313	           &   $3.3736 $   &  $21.28\pm0.3$$^e$   &   $-1.4 \pm 0.3$ &Fe+0.3$^f$     &  MgI 2852	               &VLT/Xshooter       &  (14),(15)     \\
090323 	           &   $3.5690 $   &  $20.76\pm0.08$   &   $0.25 \pm0.09$ &   Zn      &  SiII1808	               &VLT/FORS2          &   (16)    \\
090926A	           &   $2.1071 $   &  $21.60\pm0.07$   &   $-1.85\pm0.10$ &   S       &  FeII 2374	               &VLT/X-shooter      &   (17)    \\
100219A            &   $4.6672 $   &  $21.14\pm0.15$   &   $-1.1 \pm 0.2$ &   S       &  SII1 253	               &VLT/X-shooter      &   (14)    \\
111008	           &   $5.0    $   &  $22.30\pm0.06$   &   $-1.70\pm0.10$ &   S       &  NiII 1370	               &VLT/X-shooter      &   (18)    \\
120327A		   &   $2.8145 $   &  $21.01\pm0.09$   &   $-1.17\pm0.11$ &   Zn      &  CrII 2056		       &VLT/X-shooter	   &   (19)       \\
120815A	           &   $2.358  $   &  $21.95\pm0.10$   &   $-1.15\pm0.12$ &   Zn      &  MnII 2594	               &VLT/X-shooter      &   (20)    \\
\hline
\end{tabular}   
~\\\\
$^a$Transition line used to determine the velocity width of low-ionization line profiles.\\
$^b$References:
(1) \citet{Castro03}; (2) \citet{Savaglio06}; (3) \citet{Vreeswijk04}; (4) \citet{Watson06}; (5) \citet{Berger06}; 
(6) \citet{Ledoux09} ; (7) This work; (8) \citet{Piramonte08}; (9) \citet{Fynbo06};  
(10) \citet{Price07} ; (11) \citet{Eliasdottir09}; (12) \citet{Decia11}; (13) \citet{Delia11}; 
(14) \citet{Thone13} ; (15) \citet{deugraet10}; (16) \citet{Savaglio12}; (17) \citet{Delia10}; 
(18) \citet{Sparre14} ; (19) \citet{D'Elia14} ; (20) \citet{Kruhler13}
\\
$^c$ ESO Science Achive, Program Id: 075.A-0385(A), \citet{Fynbo09}
\\
$^d$ ESO Science Achive, Program Id: 080.D-0526(A) 
\\
$^e$ J. X. Prochaska, private communication.
\\
$^f$ Only iron lines are available for this system and we therefore follow the
standard procedure \citep{Rafelski12} and apply an upward
correction of $0.3$ dex to the metallicity, to correct for iron depletion and
$\alpha$-element enhancement.

\end{minipage}   
\end{table*}
Alternative selection techniques to luminosity-limited galaxy samples
allow us to form a complementary picture of the MZ evolution with
cosmic time (the MzZ relation hereafter, following
\citet{Christensen14}). Long duration Gamma-ray burst (GRB) selected galaxies are
preferentially blue, star-forming galaxies \citep[e.g.][]{LeFloch03,Christensen04} 
with low metallicities \citep[e.g.][]{Savaglio09, Graham13}. Earlier
studies were biased towards GRBs with bright optical afterglows, and
recent unbiased GRB samples \citep{Hjorth12} have shown that the
hosts follow the general trend of star-forming galaxies at similar
redshifts \citep{Michalowski12}.  Damped Lyman-$\alpha$ (DLA)
absorption line systems which arise in the GRB host galaxies can be
observed out to very high redshifts (e.g. \citet{Sparre14} analyse 
the DLA system  at $z\sim5$ in the GRB 111008A afterglow spectrum and 
\citet{Chornock14} obtain metallicity contraints of GRB 140515A at 
$z=6.33$)
and their metallicity can be measured accurately from absorption lines
which arise in the ISM. Since GRB afterglows are
transients one can study the host galaxy properties later when the GRB
afterglows have faded away.  Therefore, GRBs are ideal systems to study
the MZ relation and its evolution at high redshifts.

Conventional studies of DLA systems in quasi stellar object (QSO)
spectra are used to probe a differently selected population of high redshift 
galaxies. 
Metallicities of intervening QSO-DLAs have been measured accurately for
several hundred systems at redshifts out to $z=5$ \citep{Pettini97,
Ledoux02, Prochaska03, Rafelski12}. However, due to the
large difference in magnitudes between the bright background QSOs and
the continuum emission from the much fainter foreground galaxies,
these are extremely difficult to detect when the line of sight to the
galaxy and QSO is very close. This prevents the direct measurement of
the stellar masses for most DLA galaxies in the sightlines of QSOs,
and prevents a direct comparative MZ study of this population. To
date, only five QSO-DLA galaxies at $z>2$ have measured stellar masses
\citep{Krogager13, Fynbo13, Christensen14}.

While masses of the QSO-DLA galaxies are not known for the majority of
the population, the velocity width of low-ionisation species of
absorption lines can be used as a proxy for the mass. Indeed,
measurements of the velocity widths, defined as $\Delta v_{90}$, are
shown to correlate with the QSO-DLA metallicities \citep{Ledoux06,
Prochaska08, Moller13, Neeleman13}. In addition, \citet{Moller13} find evidence for
redshift evolution of this correlation,
reminiscent of the evolution of the MzZ relation for
luminosity-selected galaxies.  Simulations demonstrate that 
galaxies with more massive halos are more likely to produce 
metal absorption lines in the cold gas of DLA systems  
with large velocity widths, while small halos produce more of metal absorption lines 
with low velocity widths
\citep{Pontzen08,
Tescari09,Bird14} supporting the interpretation of the velocity
width-metallicity (VZ) relation as a mass-metallicity (MZ) relation
for QSO-DLAs. In what follows we shall therefore use both VZ and MZ,
chosen for clarity in the given context, to describe the relation.
Whether a VZ relation also holds true for GRB-DLA galaxies is not
yet known. \citet{Prochaska08} analyse four GRB host galaxies, and
while the four galaxies are in the VZ locus of QSO-DLAs, the small
sample size does not allow a conclusion about the existence of a VZ
relation for GRB-DLAs. 

Although both QSO-DLAs and GRB-DLAs are defined based on the large
column density of neutral hydrogen, they are selected in different ways;
GRB-DLA systems are selected based on the star formation rate (SFR) of the
galaxy hosts, while QSO-DLAs are absorption cross-section
selected galaxies \citep{Moller98, Fynbo99, Prochaska08}.
Hence, they could be drawn from distinct populations of
high-redshift galaxies and it is not known whether an MZ relation
holds for GRB-DLA systems.

The aim of this paper is to investigate the VZ correlation for a
complete literature sample of GRB-DLA galaxies, and place that into
context with the relation from QSO-DLAs. Throughout this paper,
when we refer to GRBs they are always long duration GRBs. 

The paper is organized in the following way.
The sample selection is given in section \ref{samp}. We discuss the
velocity width and the effect of spectral resolution in subsections
\ref{velocity} and \ref{correction}. In subsection \ref{final-s},  the
final sample is presented. Our results on the VZ relation for GRB-DLAs
and several aspects of comparisons between GRB host galaxies and
QSO-DLAs are presented in section \ref{are?}.  A summary of this work
and our conclusions are given in section \ref{summary}.

\section{Data}
\label{samp}
\subsection{Sample selection}
In order to compile a GRB host DLA sample suitable for comparison to the
existing samples of QSO-DLAs, we follow \citet{Moller13} and
search the literature for DLA systems in optical spectra of GRBs.
In order to be included in our sample the DLA must fullfill the
following requirements:
\\
1) $\log{N_{HI}}\geq20.3~cm^{-2}$
\\
2) it must be intrinsic to the GRB host
\\
3) it must have a reported absorption metallicity
\\
4) it must have at least one unsaturated low ionization line with
signal-to-noise appropriate for determination of velocity width.

We find a total of $20$ DLAs fulfilling all criteria, and they
make up our complete literature sample spanning the
redshift range from $z=1.97$ to $z=5.0$ (Table \ref{Tab1}).
Five of the systems have high resolution observations (VLT/UVES,
FWHM$\sim7~kms^{-1}$), $8$ medium (FWHM$\sim30-60~kms^{-1}$),
and $7$ low resolution (FWHM$\sim 110-480~kms^{-1}$). 


\subsection{Velocity width}
\label{velocity}	
To ease comparison with previous work, we use the definition of velocity
width given in \citet{Prochaska97} which is the velocity interval that
contains $90\%$ of the area under the apparent optical depth
spectrum ($\Delta v_{90}$). 
We follow \citet{Ledoux06} and \citet{Moller13} for the line selection rules and measuring method. 
The lines we select to measure $\Delta v_{90}$ for GRB-DLAs in our sample are listed in Table \ref{Tab1}.  

\begin{figure*}
\captionsetup[subfigure]{labelformat=empty}
\begin{center}$
\begin{array}{cc} \hskip -0.4 cm
\subfloat[]
{\includegraphics[trim = 0mm 0mm 0mm 0mm, clip, width=0.5\textwidth, angle=0]{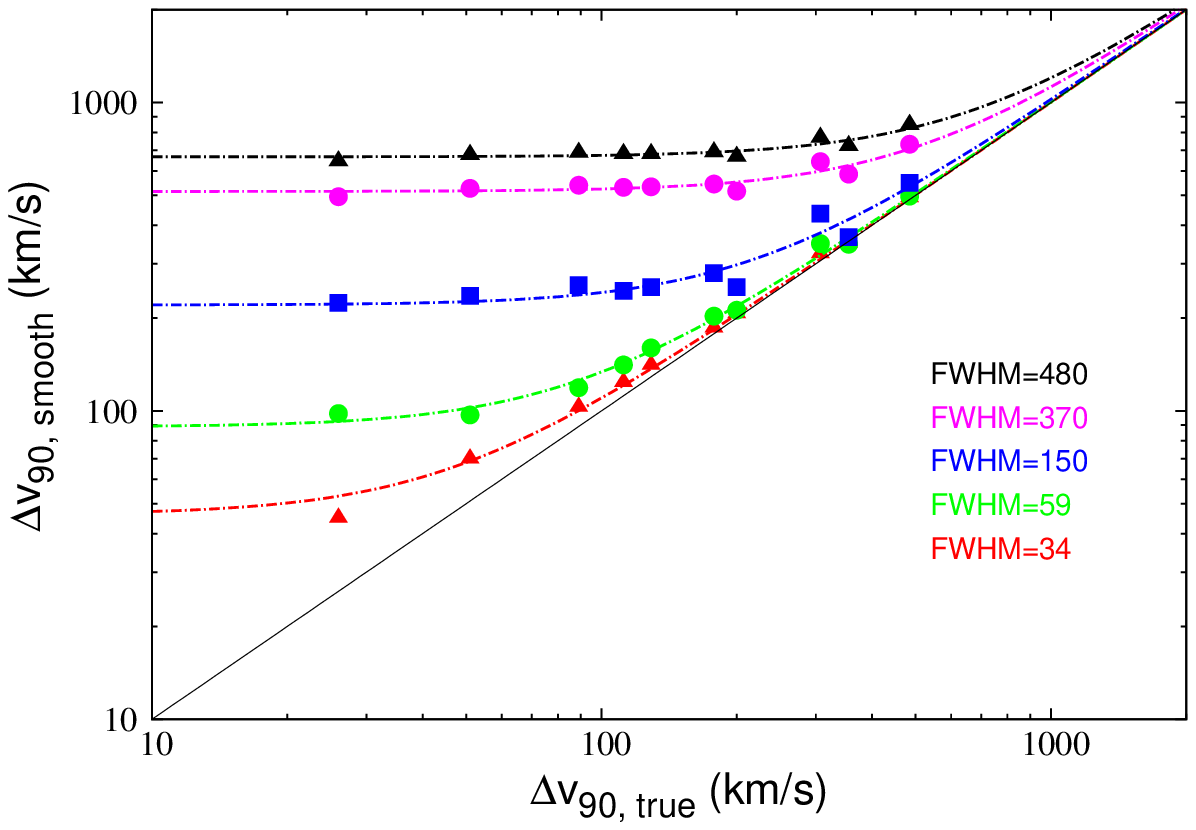}\label{sample}} &
 \hskip -0.0 cm
\subfloat[]
{\includegraphics[trim = 0mm 0mm 0mm 0mm, clip, width=0.5\textwidth, angle=0]{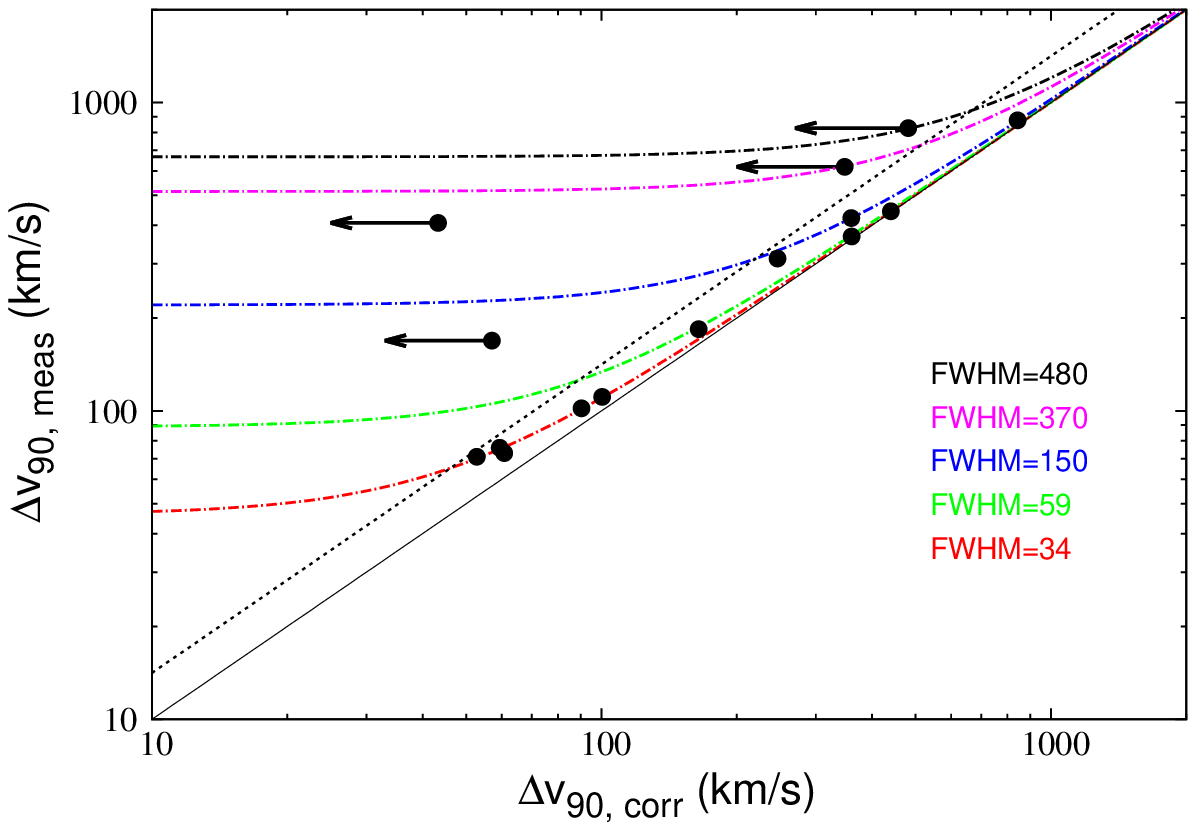}\label{example}}
\end{array}$
\end{center}
\vskip -7.7 mm
\caption{
Left panel: $\Delta v_{90}$ of the smoothed line profiles given
in Table \ref{Tab2} versus $\Delta v_{90}$ of the original profiles. 
Dash-dotted curves show the effect of smoothing to low resolution
data as explained in section \ref{velocity}. The solid black line 
marks $\Delta v_{90, smooth}$=$\Delta v_{90, true}$; 
Right panel: Dash-dotted curves and full line are the same as in 
left panel. The dotted line marks $r=0.4$ (see section \ref{correction}).
Filled circles are the measured values for medium/low resolution lines in
our sample. Four points are to the left of $r=0.4$ so we obtain
only upper limit for their intrinsic widths (see Table \ref{Tab3}).
}
\label{smooth}
\end{figure*}

\begin{table}
\begin{minipage}{83mm}
\caption{
High resolution line profiles used to correct $\Delta v_{90}$ for low/medium resolution line profiles
} \label{Tab2}
\begin{tabular}{@{}lclclc}
\hline
 &  &  & & &   \\
Quasar/GRB  & $z_{\rm DLA}$ &$\lambda$ $^a$ & $\Delta v_{90}$ & Instrument & Ref.$^b$ \\
 &  &  &  &    \\
\hline
B 2355-106 & 1.172 &  2852      & $129$ &VLT/UVES  & (1)  \\
J 1201+2117 & 3.798 &  1808    & $485$ &Keck/Hires  & (2)  \\
Q 0449-1645 & 1.007 &  1862   & $200$ &VLT/UVES  & (3)  \\
Q 0454+039 & 0.860 &  2260     & $112$ &VLT/UVES  & (4) \\
J 1107+0048 & 0.740 &  2852     & $178$ &VLT/UVES  & (5) \\
J 0256+0110 & 0.725 &  2852     & $355$ &VLT/UVES  & (5) \\
GRB 050730 & 3.969 & 1618     & $26$ &VLT/UVES  & (6) \\
GRB 050820A & 2.615 &1608$^c$    & $307$ &VLT/UVES  & (7) \\
GRB 050922C & 2.199 &  1608    & $89$ &VLT/UVES  & (8) \\
GRB 071031 & 2.692 &  1559 & $51$ &VLT/UVES  & (9) \\  
\hline
\end{tabular}   
~\\
$^a$Rest-frame wavelength of  transition lines smoothed to low/medium resolutions.\\
$^b$References:
(1) \citet{Ellison12}; (2) \citet{Rafelski12}; (3) \citet{Peroux08}; (4) \citet{Pettini00}; (5) \citet{Peroux06}; 
(6) \citet{Ledoux09}; (7) This work; (8) \citet{Piramonte08}; (9) \citet{Fox08}. \\
$^c$ ESO Science Achive, Program Id: 075.A-0385(A)
\end{minipage}   
\end{table}

The accuracy of $\Delta v_{90}$ measurements critically depends  on the resolution 
of the spectra. 
Quasar DLA absorption line analysis is effectively only carried out on
spectra with resolution high enough that the absorption line
profiles are well sampled. Due to the rapid fading of GRB optical transients (OTs), most
often we do not have the luxury of high resolution spectra, and we must
instead  use spectra of lower resolution. 
Observing at low
resolution causes a smearing of the absorption lines which will in
turn lead to a measurement of $\Delta v_{90}$ which is larger than the
true $\Delta v_{90}$. The magnitude of the effect depends on both the
true $\Delta v_{90}$ and the resolution of the spectrum. If the
true width is large then even very low resolution spectra may still be
used, if the true width is very small then the width information may have
been lost completely even in a medium resolution spectrum 
and hence only an upper limit for the velocity width can be derived.  
For each resolution there is a range of line widths where the effect of
the smearing can be computed and corrected for.  \citet{Prochaska08}
showed how such a correction could be carried out for observations 
with FWHM$=45~kms^{-1}$ resolution. Here we seek to generalize this
method for a large range of resolutions.

\subsection{Velocity width correction}
\label{correction}
First we examine our data set and identify  the resolution (FWHM) of
each of the different instrument/setting configurations used (listed in
Table~\ref{Tab3}). Next we search the literature and
select several high resolution (VLT/UVES or Keck/Hires) observed
line profiles (Table \ref{Tab2}) with $\Delta v_{90}$ covering a wide
range of widths ($\sim25-485~kms^{-1}$). 
The velocity widths of those high-resolution profiles are considered to 
be the true line widths. We then smooth  each of the high resolution lines  
to each of the
lower resolutions (to smooth a line, we convolve  it with a Gaussian with $\sigma$ 
related to the lower resolution) 
and measure  the 
$\Delta v_{90,{\rm smooth}}$. 
In Fig.~\ref{smooth} (left panel) we plot the resulting widths of the
smoothed lines versus their true widths for 5 representative resolutions, 
using different symbols (and colour coded in the online version) for each
resolution. We find that a simple hyperbola of the form
\begin{eqnarray}
\label{hyper}
\Delta v_{90,{\rm smooth}}=(\Delta v_{90,{\rm true}}^2+a^2)^{0.5}
\end{eqnarray} 
where $a=1.40 \times {\rm FWHM}$, provides an excellent fit in all cases.
The best fit curves of the form given in equation \ref{hyper} (determined by using a
nonlinear least-squares algorithm) are shown in Figure \ref{smooth}.
We can now use those fitted curves as a prescription to correct our
medium and low resolution data back to their intrinsic values.
\\
As pointed out above, the correction can only be trusted if the
resolution FWHM is not large enough to completely dominate over the
intrinsic width of the line. In case the intrinsic width is equal to
the resolution then the measured width will be a factor $\approx \sqrt 2$
larger, which is fully possible to correct for. We have therefore chosen a
conservative approach to only trust corrections if the measured width is
less than 1.4 times the width after correction. In other words, we
define a parameter $r$ such that
\begin{eqnarray}
\label{r}
r:=\frac{\Delta v_{90, meas}-\Delta v_{90, corr}}{\Delta v_{90, corr}}.
\end{eqnarray} 
and only consider systems correctable if $r \leq 0.4$. The line
corresponding to $r=0.4$ is shown as a dotted  straight line in Figure \ref{smooth}
(right hand panel). It is seen that four systems are to the left side of the $r=0.4$ line, 
on the flat, uncorrectable part of the curves. These four systems are consequently excluded from further analysis when  
$\Delta v_{90}$ is involved.

\subsection{Final sample}
\label{final-s}
In Table \ref{Tab3} we provide $\Delta v_{90}$ before and after correction, and also $r$ values for the $20$ DLAs in our sample. 
For four DLA systems (GRBs 030323, 050401, 050505
and 070802)  we cannot reliably correct
$\Delta v_{90}$ and do not include 
them in any further analysis which involve  $\Delta v_{90}$. Our final sample consists of $16$ GRB host galaxies, spanning the
redshift range from $z=1.97$ to $z=5.0$.

\begin{table}
\begin{minipage}{79mm}
\caption{
Final results for velocity widths of the GRB-DLA sample
} 
\label{Tab3}
\begin{tabular}{@{}lcccc}
\hline
 &  & &   & \\
GRB  & $\Delta v_{90, meas}$  &$\Delta v_{90, corr} $ & $r$  & line resolution	  \\
  &$(kms^{-1})$ & $(kms^{-1})$ & & $(kms^{-1})$      \\
&&&&\\
\hline
\multicolumn{5}{c}{High Resolution} \\
\hline
050730	 &$    34     $&$    34     $&$     ...      $&$    ...    $   \\    
050820A	 &$    300    $&$    300    $&$     ...      $&$    ...    $   \\  
050922C	 &$    89     $&$    89     $&$     ...      $&$    ...    $    \\
071031	 &$    86     $&$    86     $&$     ...      $&$    ...    $   \\
081008	 &$    60     $&$    60     $&$     ...      $&$    ...    $   \\
\hline                                              
\multicolumn{5}{c}{Medium Resolution}                \\
\hline                                              
000926	 &$    368    $&$    362    $&$     0.022    $&$    54     $    \\
060206	 &$    444    $&$    441    $&$     0.008    $&$    39     $    \\
090313	 &$    184    $&$    165    $&$     0.118    $&$    59     $    \\	      
090926A	 &$    71     $&$    53     $&$     0.344    $&$    34     $   \\    
100219A  &$    76     $&$    59     $&$     0.280    $&$    34     $   \\  	      
111008	 &$    111    $&$    100    $&$     0.106    $&$    34     $    \\
120327A  &$    102    $&$    90     $&$     0.130    $&$    34     $    \\	      
120815A	 &$    73     $&$    61     $&$     0.202    $&$    29     $   \\
\hline                                              
\multicolumn{5}{c}{Low Resolution} \\               
\hline			                            
030323	 &$    169    $&$    \leq57 $&$     1.963    $&$    114    $ \\	      
050401	 &$ 619  $&$    \leq348$&$     0.780    $&$    367    $    \\	      
050505	 &$ 407  $&$    \leq43 $&$     8.398    $&$    290    $   \\	            
060510B  &$    422    $&$    360    $&$     0.173    $&$    158    $    \\    
070802	 &$    826    $&$    \leq481$&$     0.716    $&$    481    $    \\	          
080210	 &$    312    $&$    247    $&$     0.265    $&$    137    $    \\	         	      
090323 	 &$    876    $&$    843    $&$     0.039    $&$    170    $    \\	         
\hline
\end{tabular}   
~\\  
The second and third columns are $\Delta v_{90}$ before and after correction respectively 
(see equation \ref{hyper}). 
$r$ values as defined in equation \ref{r} are given in the fourth column. The resolution 
(FWHM) of the selected line profiles  used for measuring velocity $\Delta v_{90}$ are given 
in the fifth column. 
For four DLA systems (GRBs 030323, 050401, 050505, and 070820) $r>0.4$ and hence 
values of $\Delta v_{90,corr}$ for these systems are upper limits.  
\end{minipage}    
\end{table}

\section{Are GRB host galaxies and DLA galaxies the same?}
\label{are?}

\subsection{Mass-Metallicity relation}

DLAs are most commonly observed in the spectra of QSOs, and a large body
of data is available in the literature for such DLAs. In this section
we investigate whether there is any evidence that the MZ relation of the
GRB host galaxies differs from that of QSO-DLAs. For that purpose, we
compare the VZ data of GRB hosts presented in section~\ref{samp} to
the QSO-DLA data presented by \citet{Moller13}.

In the upper panel of Figure~\ref{z-corrected} we present a plot of the
two data sets.  It can be seen that the
observed metallicities fully overlap.  A least-square fit of a
straight line to both of the data sets individually, shows that the
parameters of the fits are identical to within the uncertainties. For the
further analysis, we adopt the slope of 1.46 for the line, as found by
\citet{Ledoux06}, and we fit only intercept for the two samples
individually. These fits are shown as full line (GRB hosts) and dotted
line (QSO-DLAs) in Figure~\ref{z-corrected}.

\subsection{Redshift evolution}

The MZ relation of QSO-DLA galaxies, as well as that of emission line
galaxies, evolves with redshift. Our current sample is too small to
independently derive a model for the redshift evolution. However, it is
possible to investigate whether the data is consistent with any
previously proposed model for the redshift evolution (MzZ).  For each MzZ
model we use
\begin{eqnarray}\label{evol} 
{\rm [X/H]}=  {\rm [X/H]}_e + f(z)
\end{eqnarray} 
to compute an evolution corrected metallicity (${\rm [X/H]_e}$), where
$f(z)$ is the shift in the MZ relation as a function of redshift and
$z$ is the redshift of the galaxy.

We consider three different evolution models. \citet{Neeleman13} study a
sample of DLAs covering the redshift interval from $2$ to $5$, and
report that the MZ relation evolves linearly with a slope of $-0.32$.
Our first model is to use a line with this slope for the whole redshift
range covered by our data as $f(z)$. 

\begin{figure}
\captionsetup[subfigure]{labelformat=empty}
\begin{center}$
\begin{array}{c}
\subfloat[]
{\includegraphics[trim = 0mm 10mm -7mm 0mm, clip, width=0.5\textwidth, angle=0]{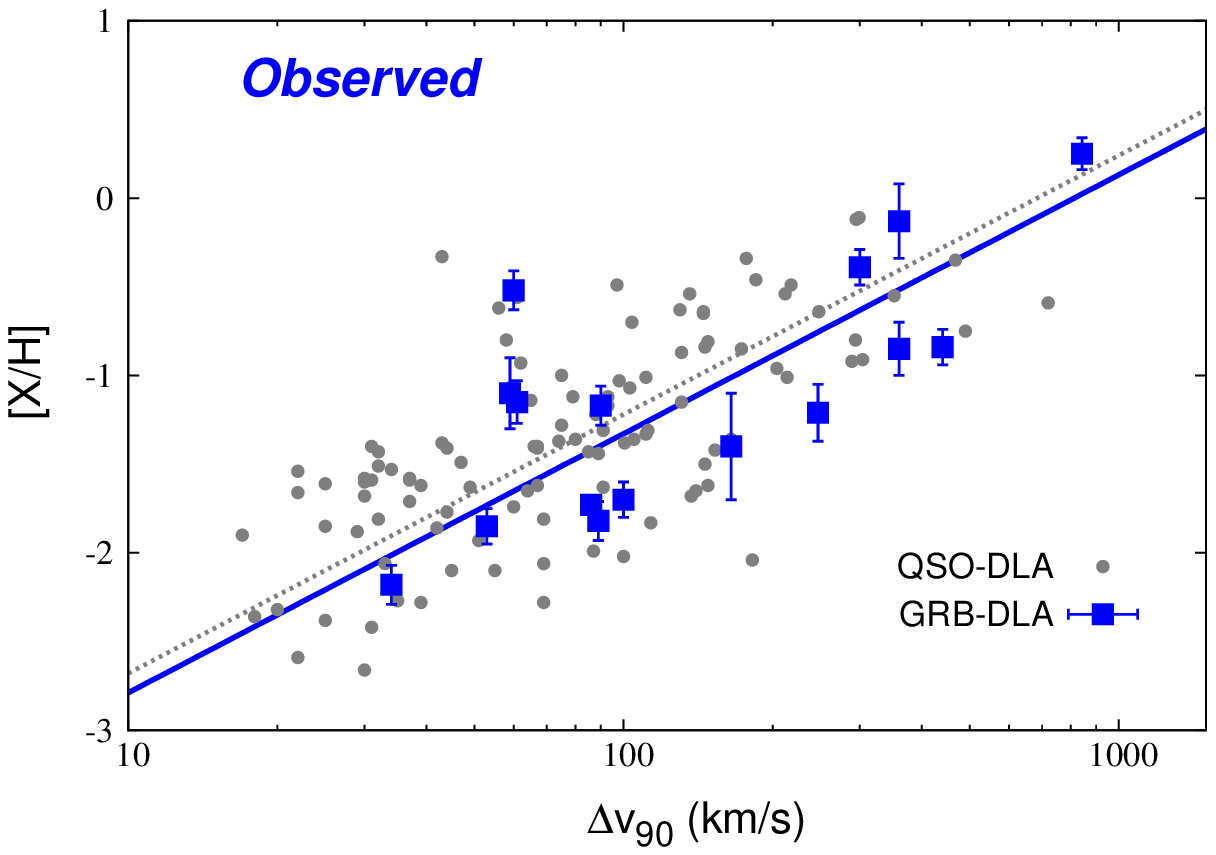}} 
\\
\subfloat[]
{\includegraphics[trim = 0mm 0mm -7mm 0mm, clip, width=0.5\textwidth, angle=0]{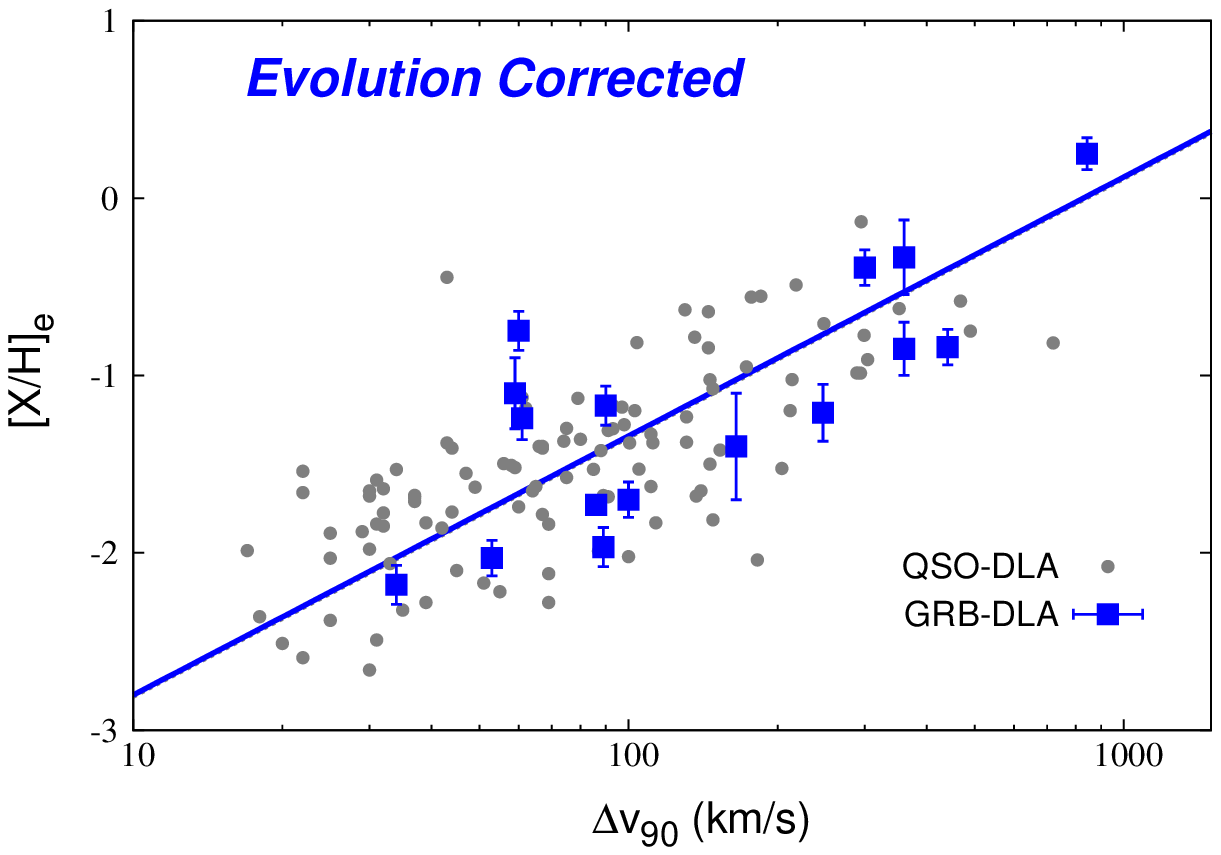}}
\end{array}$
\end{center}
\vskip -7.7 mm
\caption{Metallicity vs velocity width for QSO-DLAs (small dots) and
GRB host DLAs (large blue squares). Upper panel: observed values;
Lower panel: [X/H] values corrected for the redshift evolution
determined for QSO-DLAs. For both samples the scatter is reduced
indicating that their MZ relations follow similar redshift evolutions. 
Best fits (using equation \ref{C2}) are shown as dotted lines (QSO-DLAs), 
full lines (GRB-DLAs).  
}
\label{z-corrected}
\end{figure}

Based on a study of QSO-DLAs covering the wider redshift interval
from $0.1$ to $5.1$, \citet{Moller13} find that the MzZ relation of DLAs
cannot be well described by a single slope. Rather the relation is
constant from the highest observed redshifts to a redshift of $2.6$,
and evolves then linearly with a slope of $-0.35$ at lower redshifts.
We use this prescription
as our second model, and refer to it hereafter as the ``late evolution'' model.
In the lower panel of Figure~\ref{z-corrected},  we show the evolution
corrected VZ relation based on the ``late evolution'' model for both the GRB host
and the QSO-DLA samples.  The  scatter for the evolution corrected relation is 
lower than that of the uncorrected relation shown in the upper panel.
This is a preliminary indication that  the GRB host  data
is consistent with this evolution model. 

Finally, \cite{Maiolino08} investigate the metallicity evolution of  emission
line selected galaxies. Their data are consistent
with linear evolution with a slope of $-0.35$.  Our third evolution model is
again to adopt this slope for evolution throughout our redshift range. 

In order to compare the different evolution models, we adopt the following
procedure.  First, we determine the intrinsic scatter of the VZ  relation before
and after correcting with each of the evolution models.
Following \citet{Moller13}, we separate the total observed scatter $\sigma_{\rm
tot}$ into contributions from measurement errors $\sigma_{\rm met}(i)$
for each GRB host $i$ 
and the intrinsic scatter of the relation $\sigma_{\rm scatter}$.  For this we define 
$C_{\rm dof}^2$ as 
\begin{eqnarray} \label{C2} C_{\rm dof}^2=\sum_{i=1}^{16} (({\rm [M/H]}(i)-1.46
        && \hskip -6 mm \log{(\Delta v_{90}(i))}-zp)\cr &&\hskip 5
        mm/\sigma_{\rm tot}(i))^2/{\rm dof}, 
\end{eqnarray} 
where $\sigma_{\rm tot}(i) =\left (\sigma_{\rm met}(i)^2+\sigma_{\rm
scatter}^2\right)^{1/2}$, $zp$ is the intercept of the fitted line, and dof is the
degrees of  freedom, which in this case is 15.  We then set $C_{\rm dof}$ to its
expected value of unity, and numerically solve equation~(\ref{C2}) for
$\sigma_{\rm scatter}$ as a function of $zp$. Finally, we adopt the minimum of this 
function as the  value of $\sigma_{\rm scatter}$.

The values found for  $\sigma_{\rm scatter}$ for each of the evolution models
are reported in table~\ref{Tablast}. 
The ``late evolution'' model  provides the least intrinsic scatter, and
therefore the best fit to the data. From table~\ref{Tablast} it is
also seen that both ``constant slope'' models provide larger scatter
than the ``uncorrected'' data.  We carry  out sets of
Monte-Carlo simulations to test the significance of this finding.  In each set
of simulations, we assume one of the evolution models to be true, and then
count the number of times that the  intrinsic scatter of the evolution
corrected MZ relations computed with different models behave as the real data.

\begin{table}
\caption{
Redshift evolution models and their effect on the intrinsic scatter of the VZ relation 
} 
\begin{center}
\begin{tabular}{lcc}
\hline
 &  &    \\
 Redshift evolution model & Observed $\sigma_{\rm scatter}$ & Relative likelihood$^a$ \\
 &  &    \\
\hline
late evolution             & $0.411$ & $1$    \\ 
Constant slope of $-0.32$  & $0.453$ & $0.060$    \\
Constant slope of $-0.35$  & $0.467$ & $0.035$    \\
Uncorrected                & $0.446$ & $0.255$      \\
\hline
\end{tabular}   
\end{center}
\noindent$^a$ Relative number of times that the simulation of a model reproduces the observed behaviors in scatter.
\label{Tablast}
\end{table}

The detailed procedure is the following. In each of the simulations, we
assign to each GRB host a value for [X/H]$_e$ based on its redshift using
one of the evolution models, and its measured $\Delta v_{90}$. We then add
normally distributed noise based on the measurement errors $\sigma_{\rm met}$,
and additional intrinsic noise $\sigma_{\rm scatter}$.
We then compute the observed evolved metallicities with each of the
evolution models, and determine $zp$ and 
intrinsic scatter as we have done for the real data.  For each
imposed evolution model, we perform $10^6$ simulations.  We then  count the
number of times that applying each of the evolution corrections moves the
intrinsic scatter as much as the real data do or more. 

The relative number of times this happens is reported in
table~\ref{Tablast}.  We conclude that if the  ``late evolution'' model is
true, it is more than 17 times as likely to observe the scatter behave as
observed than if any of the fixed slope evolution models are true. 
In conclusion of this section, we find that 
among the 3 models tested,  
GRB hosts are in better agreement with the late evolution model with a break around $z\sim2.6$.

\subsection{GRB host metallicities}
\label{histos}
\begin{figure}
\includegraphics[width=84mm]{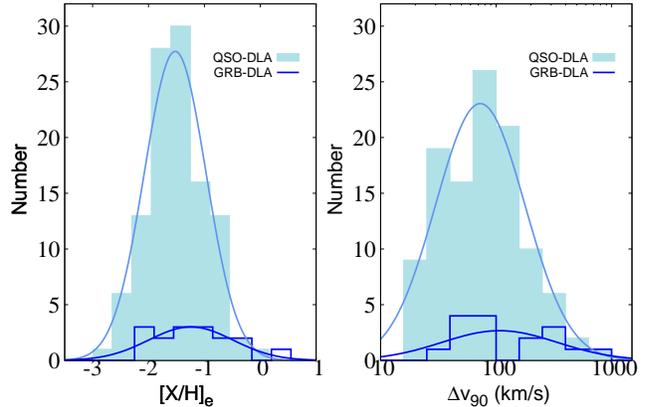}
\vskip -3 mm
\caption{Histograms of metallicity and $\Delta v_{90}$ are presented in left and right panels respectively. 
The light blue shade show the histograms for the 110 QSO-DLA galaxies \citep{Moller13} and the solid 
dark blue lines are for the 16 GRB-DLA galaxies in our sample (see Tables \ref{Tab1} and \ref{Tab3}). 
Best fit Gaussians are overplotted.
}
\label{vz-dist}
\end{figure}

It has been pointed out that GRB hosts in general have higher observed
metallicities than QSO-DLAs at similar redshifts \citep{Fynbo06,
Savaglio06, prochaska07a, Fynbo08, Cucchiara14}. 
In Figure \ref{vz-dist} (left panel) we show the histogram of metallicities
(corrected for redshift evolution, i.e. the projection onto the left
axis of Figure \ref{z-corrected}, lower panel) for both GRB hosts
and QSO-DLAs. The median metallicity for the two samples is
$-1.19$ and $-1.53$ respectively, consistent with previous reports.
In the right panel we show the corresponding histograms
for the $\Delta v_{90}$, and it is seen that there also is a
corresponding shift of the GRB hosts towards slightly larger
velocity widths (median of $95$ km/s versus $75$ km/s for QSO-DLAs).
Seen together the two shifts thus form a shift of the GRB hosts along
the relation towards the upper right such that both samples follow the
same relation, but the GRB hosts populate the part of the diagram for
slightly larger masses than QSO-DLAs. One possible way of understanding 
this could be that QSO-DLAs and GRB hosts are selected in two different
ways from the same underlying sample. Following \citet{Fynbo08} one
may argue that GRB hosts are selected by SFR (as already demonstrated by
\citet{Christensen04}), i.e. weighted $\propto L$ (luminosity), while
QSO-DLAs are selected by cold gas projected absorption area, i.e.
$\propto R^2$ (gas disk radius squared). We know that $R^2 \propto L^{2t}$
where $t$ is the Holmberg parameter \citep{Fynbo99}, and it is now
easy to understand how a shift along the relation may occur. For
$t=0.5$ there will be no shift, for $t < 0.5$ QSO-DLAs will preferentially
be found to the lower left relative to GRB hosts, while for $t > 0.5$ they
will in general be more Luminous than GRB hosts and therefore be found in
the upper right. From Fynbo etal, (1999) we see that $t$ is $0.4$ at $z=0$
but that it was smaller ($0.25$) in the past (at higher redshifts). As a
consequence we predict that GRB hosts, on average, will be slightly
more massive than QSO-DLAs and that the difference will be larger at
higher redshifts. This may be the reason for the metallicity offset 
reported, but there
is an additional effect which may cause GRB hosts to be shifted in
the lower plot of Figure \ref{z-corrected}. The effect is related to the
different impact paramet distributions, it is discussed in the following
sections and shown in Figure \ref{zb-corrected}. Note that where the
effect of selection bias discussed above
should only shift along the relation, the effect discussed below is more
complex and may possibly result in a shift away from the relation. It is
therefore not trivial that the two relations match so well in
Figure \ref{z-corrected}, they could have formed two separate relations.

\subsection{Impact parameter, metallicity gradient and gravitational well}

The one thing we know is different between QSO-DLAs and GRB host DLAs
are the sightlines. A QSO-DLA is sampling the HI gas in the intervening
DLA galaxy and its halo via random selection. This results in a
distribution of impact parameters reflecting the size, shape and
inclination of the gas associated with the galaxy
\citep[for details see discussion in][]{Moller98}. In contrast GRBs
are mostly located closer to the centres of their hosts (typically
few $kpc$), and therefore they sample gas closer to the centre of
the galaxy.

\begin{figure}
\includegraphics[width=87mm]{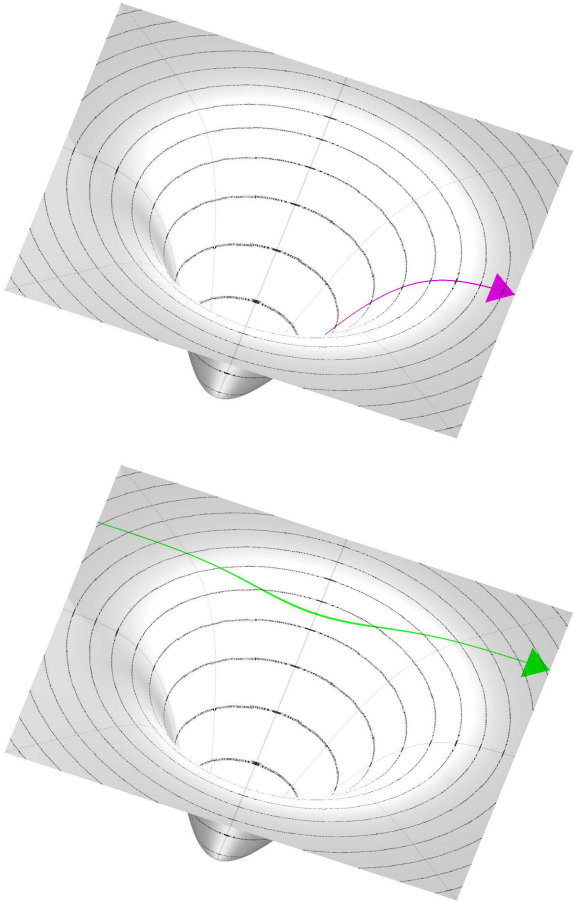}
\caption{
Paths of light through the potential well of the host galaxy. 
The upper panel illustrates a sightline from a GRB that explodes in the centre 
of the host galaxy, while the lower panel illustrates the random line of sight 
of a QSO that intersects an intervening galaxy.
}
\label{cool}
\end{figure}

This difference in paths changes the observed spectral lines in 
three different ways.
First the HI column density of GRB-DLAs
is higher because of the known impact parameter vs HI column density anti
correlation \citep{Moller98,Zwaan05,Krogager12},
second this will have an effect on the measured metallicity if the
galaxies have metallicity gradients \citep{vanzee98, Swinbank12},
and third the sightlines will sample different
paths through the dark matter gravitational well of the galaxy and
they will therefore sample different depths of this gravitational
well. We illustrate this in Figure \ref{cool}. In the bottom panel we see a
typical QSO-DLA sightline through the shallow part of the gravitational well, in the top
panel the light from a GRB passes from the centre through the deep
part of the gravitational well, but only through half of it.

The first point listed above has no effect on our observations, but
the other two will move the data points in the VZ
plots as explained below.

\subsubsection{Metallicity gradient}

For easy comparison we correct measured metallicities to the
metallicity at the centre of the galaxy. The metallicity gradient of
DLA galaxies has been determined observationally to be $-0.022 \pm 0.004$
dex per kpc \citep{Christensen14}, but for QSO-DLAs the impact
parameter is mostly unknown, and the authors also give the
observationally determined mean correction which is $0.44 \pm 0.10$.
In Figure \ref{zb-corrected} we show this "corrected to central" mass
metallicity relation for QSO-DLAs.

\begin{figure}
\includegraphics[width=87mm]{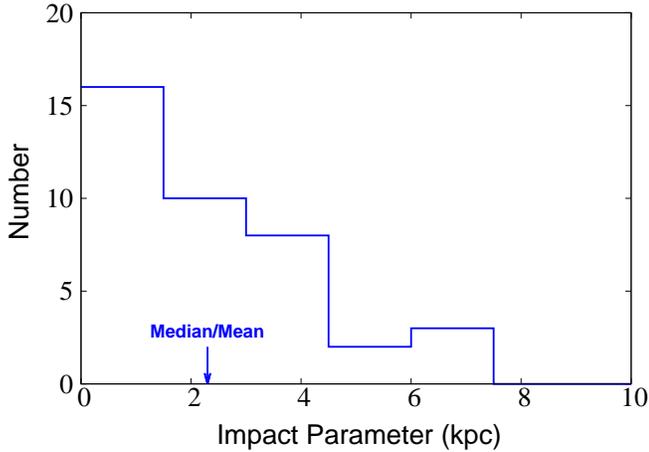}
\vskip -3 mm
\caption{
Histogram of impact parameters of GRB host galaxies given in
Table \ref{Tab4} with both mean and median equal to $2.3~kpc$  
}
\label{b-histo}
\end{figure}

We search the literature for measurements of impact parameters of
the GRBs in our sample but find that only two have been reported
\citep{Castro03,Thone13}. We find that the image of the
host has been obtained for an additional GRB \citep{D'Elia14} 
for which we measure the impact parameter.
All three values can be found in Table \ref{Tab4}.  
For the impact parameters of the remaining sample we shall use a mean value determined from a representative sample of GRB hosts.  
Such a sample is provided in \citet{Bloom02} but we find that for a
number of the OTs in that sample better astrometry was subsequently
provided by \citet{Fruchter06}, and for those we  reassess  the impact
parameters based on the \citet{Fruchter06} data.
From the sample of \citet{Perley13} we include all GRBs for which 
coordinates of both hosts and OTs are provided with uncertainty
$\leq 0.3 ''$, and for which redshifts are known.
The final values of impact parameters are presented in Table \ref{Tab4} and shown in histogram form in Fig \ref{b-histo}.  

Based on this table we find that the mean, the weighted mean, and
the median values are $2.3~kpc$, $2.5~kpc$, and $2.3~kpc$ respectively.
All those values are small and very similar and we choose to use the
mean value for those hosts with no measured impact parameter. Following
\citet{Christensen14}, we use the metallicity gradient of $-0.022$ dex per $kpc$ and correct the metallicity of the 
GRB-DLAs in our sample to the central metallicity (Figure \ref{zb-corrected}).

\begin{table}
\begin{minipage}{79mm}
\caption{
List of impact parameters 
} 
\label{Tab4}
\begin{tabular}{@{}lcccc}
\hline
 &  & &   & \\
GRB  & Redshift  & impact parameter& error-bar  & Reference$^a$	  \\
  && $(kpc)$ & $(kpc)$ &       \\
&&&&\\
\hline                                              
\multicolumn{5}{c}{GRBs in our sample}                \\
\hline 
000926          &  $    2.038   $&  $  0.29     $&  $   0.037	$& (1)   \\
100219A		&  $    4.667   $&  $  2.6      $&  $   0.2	$& (2)   \\
120327A 	&  $    2.815   $&  $  3.19     $&  $   2.39	$& (3)    \\
\hline
\multicolumn{5}{c}{other GRBs}                \\
\hline 
970228		&  $	0.695 	$&  $  0.17 	$&  $	0.24	$& (3) 	\\	  	 	
970508	        &  $	0.835   $&  $  0.09     $&  $   0.09	$& (4)    \\
970828	        &  $	0.958   $&  $  4.05     $&  $   4.33	$& (4)    \\
971214 	        &  $	3.418   $&  $  1.11     $&  $   0.56	$& (4)    \\
980425 	        &  $	0.008   $&  $  2.34     $&  $   0.01	$& (4)    \\
980613 	        &  $	1.096   $&  $  0.78     $&  $   0.67	$& (4)    \\
980703A	        &  $	0.966   $&  $  0.96     $&  $   0.54	$& (4)    \\
990123 	        &  $	1.6     $&  $  6.11     $&  $   0.03	$& (4)    \\
990506 	        &  $	1.31    $&  $  2.68     $&  $   4.14	$& (4)    \\
990510 	        &  $	1.619   $&  $  0.60     $&  $   0.08	$& (4)    \\
990705 	        &  $	0.84    $&  $  7.17     $&  $   0.38	$& (4)    \\
990712 	        &  $	0.434   $&  $  0.30     $&  $   0.49	$& (4)    \\
991208 	        &  $	0.706   $&  $  0.00     $&  $   0.62	$& (3)    \\
991216 	        &  $	1.02    $&  $  3.12     $&  $   0.28	$& (4)    \\
000301C	        &  $	2.03    $&  $  0.62     $&  $   0.06	$& (4)    \\
000418          &  $    1.118   $&  $  0.20     $&  $   0.56	$& (4)    \\
010222          &  $    1.477   $&  $  0.376    $&  $   0.76	$& (3)    \\
010921          &  $    0.45    $&  $  2.53     $&  $   0.52	$& (3)    \\
011121          &  $    0.362   $&  $  3.98     $&  $   1.11	$& (3)    \\
011211          &  $    2.141   $&  $  2.97     $&  $   2.97	$& (3)    \\
020405          &  $    0.69    $&  $  5.95     $&  $   1.57    $& (3)    \\
020813          &  $    1.255   $&  $  0.18     $&  $   0.19    $& (3)    \\
020903          &  $    0.251   $&  $  2.23     $&  $   2.23    $& (3)    \\
021004          &  $    2.330   $&  $  0.37     $&  $   0.37  	$& (3)    \\
021211          &  $    1.006   $&  $  0.71     $&  $   0.36    $& (3)    \\
030115          &  $    2.5     $&  $  2.53     $&  $   1.52    $& (3)    \\
030329          &  $    0.168   $&  $  0.37     $&  $   0.13    $& (3)    \\
040924          &  $    0.859   $&  $  1.70     $&  $   1.02	$& (3)    \\
041006  	&  $    0.716   $&  $  2.54     $&  $   1.28	$& (3)	\\
050915A         &  $    2.527   $&  $  6.72     $&  $   0.90	$& (5)   \\
051022          &  $    0.8    $&   $  1.83     $&  $   1.51	$& (5)   \\
061222A         &  $    2.088   $&  $  3.12     $&  $   1.03	$& (5)   \\
070802          &  $    2.455   $&  $  3.63     $&  $   1.50	$& (5)   \\
080325          &  $    1.78    $&  $  5.73     $&  $   1.04	$& (5)   \\
080607          &  $    3.036   $&  $  3.30     $&  $   1.43	$& (5)   \\
081221          &  $    2.26    $&  $  3.97     $&  $   0.76    $& (5)   \\
\hline                                                                          
\end{tabular}   
~\\
Redshift of GRBs are given in second column.   
The third and fourth columns present the impact parameter and
corresponding error. \\
$^a$ References: (1) \citet{Castro03}; (2) \citet{Thone13}; (3) this work; (4) \citet{Bloom02}; (5) \citet{Perley13} 
\end{minipage}    
\end{table}

\subsubsection{Potential well depth}
The line of sight to the OT in a GRB host samples only half
of the gravitational well in which the host resides (see Figure \ref{cool}, upper
panel). It is hard to predict exactly which value for $\Delta v_{90}$ one
would have observed in case the line had been complete through the
other side of the host, but it certainly would not be less than for
the observed half galaxy. One could make the simple assumption that
the cold clumps in the ISM and halo gas move randomly, but that most
are bound inside the gravitational well. In that case $\Delta v_{90}$ would
increase by about a factor $\sqrt 2$.  The same type of argument
can be applied to sightlines to QSOs through intervening DLA galaxies.
Such sightlines will in most cases not pass close to the centre, but
rather at impact parameters of order $\approx 10$kpc (see Figure \ref{cool}, lower
panel). Again, only a fraction of the full well will be sampled, and
the measured $\Delta v_{90}$ will be smaller than if sampled through the
centre. The precise magnitude of those effects is hard to quantify.
Detailed high resolution hydro-dynamic simulations would be required to
get an estimate.

\begin{figure}
\includegraphics[width=87mm]{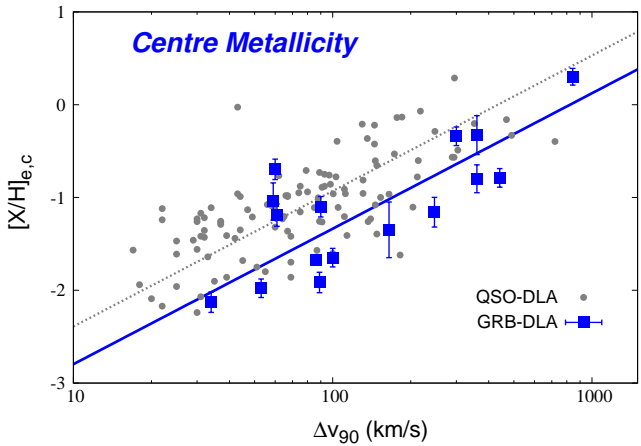}
\vskip -3 mm
\caption{ 
Metallicity corrected with the late evolution model  and also 
corrected to the central metallicity ($[X/H]_{e,c}$) 
vs velocity width for QSO-DLAs (small dots) and
GRB host DLAs (large blue squares). 
Best fits (using equation \ref{C2}) are shown as dotted lines (QSO-DLAs), 
full lines (GRB-DLAs).  
}
\label{zb-corrected}
\end{figure}

In Figure \ref{zb-corrected} we apply the empirical corrections for metallicity
gradients, and it is seen that there is now a large shift between the
two populations. If we interpret this shift as an effect of different
sampling of gravitational wells, then the shift corresponds to a
factor $2$ change of  $\Delta v_{90}$  for galaxies with the same metallicity at a
given redshift. The shift is in the sense that QSO-DLAs have  $\Delta v_{90}$ 
a factor of $2$ less than a GRB host with the same metallicity. The shift
is therefore in the direction we would expect in case it is due to the
gravitational well sampling effect. This may therefore be the effect we
have predicted, but at present it is not possible to conclusively
prove this.

It is curious, though, that the two samples overlap so perfectly in Figure \ref{z-corrected}.  
This means that either there are no metallicity gradients in neither
QSO-DLA galaxies, nor in GRB hosts, and also there is no effect due to
the gravitational well sampling - or,
the effect of metallicity gradients exactly cancels the effect of
gravitational well sampling.  If the latter is the case then this
means that the general concept of an MZ relation plus metallicity
gradients simply is a convolved and roundabout way of describing a
much simpler underlying relation between metallicity and gravitational
well depth.

\begin{table*}
\begin{minipage}{125mm}
\caption{
For the sample of the GRB-DLA galaxies, 
redshift, impact parameter,  and metallicity are presented in second, third, and fourth columns respectively. 
Predicted and measured stellar masses are given in the fifth and sixth columns.   
} 
\label{Tab5}
\begin{tabular}{@{}lclcccccccc}
\hline
 &  &  & & &  & &&&  \\
GRB && redshift& & $b~(kpc)$&& $[X/H]$& &Predicted mass,& & Measured mass, \\
 &  &&  & & &&& $\log{(M_{\star}/M_\odot)}$ & & $\log{(M_{\star}/M_\odot)}$  \\
 &  &&  & & && & &    \\
\hline
000926    &&   $2.0379 $ &&   $0.29 $  &&  $-0.13\pm0.21$   &&   $ 9.91\pm0.81$&&   $9.52\pm0.84 $ 		\\
030323    &&   $3.3718 $ &&   $...  $  &&  $-1.26\pm0.20$   &&   $ 8.34\pm0.80$&&   $...         $ 		\\
050401    &&   $2.8992 $ &&   $...  $  &&  $-1.0 \pm0.4 $   &&   $ 8.80\pm1.01$&&   $...         $ 		\\
050505    &&   $4.2748 $ &&   $...  $  &&  $-1.2 \pm... $   &&   $ 8.45\pm>0.72$&&   $...         $ 		\\   
050730    &&   $3.9686 $ &&   $...  $  &&  $-2.18\pm0.11$   &&   $ 6.73\pm0.75$&&   $...         $ 		\\
050820A   &&   $2.6147 $ &&   $...  $  &&  $-0.39\pm0.10$   &&   $ 9.88\pm0.74$&&   $8.64^{+0.58}_{-0.23}  $ \\
050922C   &&   $2.1992 $ &&   $...  $  &&  $-1.82\pm0.11$   &&   $ 7.11\pm0.75$&&   $...         $ 		\\
060206    &&   $4.048  $ &&   $...  $  &&  $-0.84\pm0.10$   &&   $ 9.08\pm0.74$&&   $...         $ 		\\
060510B   &&   $4.941  $ &&   $...  $  &&  $-0.85\pm0.15$   &&   $ 9.07\pm0.77$&&   $...         $ 		\\
070802    &&   $2.4549 $ &&   $3.63 $  &&  $-0.50\pm0.68$   &&   $ 9.64\pm1.40$&&   $9.7^{+0.2}_{-0.3} $ 	\\
071031    &&   $2.6922 $ &&   $...  $  &&  $-1.73\pm0.05$   &&   $ 7.52\pm0.73$&&   $...         $ 		\\
080210    &&   $2.641  $ &&   $...  $  &&  $-1.21\pm0.16$   &&   $ 8.43\pm0.78$&&   $...         $ 		\\
081008    &&   $1.9683 $ &&   $...  $  &&  $-0.52\pm0.11$   &&   $ 9.26\pm0.75$&&   $...         $ 		\\
090313    &&   $3.3736 $ &&   $...  $  &&  $-1.40\pm0.30$   &&   $ 8.10\pm0.90$&&   $...         $ 		\\
090323    &&   $3.569  $ &&   $...  $  &&  $ 0.25\pm0.09$   &&   $11.00\pm0.74$&&   $11.20\pm0.75$ 		\\
090926A   &&   $2.1071 $ &&   $...  $  &&  $-1.85\pm0.10$   &&   $ 7.00\pm0.74$&&   $...         $ 		\\
100219A   &&   $4.6672 $ &&   $2.6  $  &&  $-1.10\pm0.20$   &&   $ 8.64\pm0.80$&&   $...         $ 		\\
111008    &&   $5.0    $ &&   $...  $  &&  $-1.70\pm0.10$   &&   $ 7.57\pm0.74$&&   $...         $ 		\\
120327A   &&   $2.8145 $ &&   $3.19 $  &&  $-1.17\pm0.11$   &&   $ 8.54\pm0.75$&&   $...         $ 		\\
120815A   &&   $2.358  $ &&   $...  $  &&  $-1.15\pm0.12$   &&   $ 8.39\pm0.75$&&   $...         $ 		\\
\hline
\end{tabular}   
~\\\\
References for measured stellar masses: \citet{Savaglio09} for 000926; \citet{Chen09} for 050820A; \citet{Kruhler11} for 070802; This work for 090323. 
\end{minipage}   
\end{table*}

\subsection{Stellar mass}
\label{mass}
A prescription for computation of the stellar mass of QSO-DLA galaxies
from only metallicity and redshift was given in \citet{Moller13}.
\citet{Christensen14}, 
improved this prescription by adding the effect of metallicity
gradients and also performed a 
test comparing the computed stellar mass to the measured stellar 
mass from the SED fits. This test was carried out using the complete set 
of QSO-DLA galaxies for which the test is currently possible. 
They concluded that the prescription is confirmed for galaxies of stellar 
masses down to $\log{(M_{\star}/M_\odot)}=8$, while for lower stellar 
masses there are no available data.
Here we use the prescription from \citet{Christensen14} (their equation
(3) including the metallicity gradient term $\Gamma b$) to compute the
predicted stellar masses of all the host galaxies in our sample (listed
in Table \ref{Tab5}). For 3 of those, stellar masses have
been determined directly via SED fitting (also provided in
Table \ref{Tab5}). For the host of GRB 090323, we use the photometric
data given in \citet{McBreen10} and determine the stellar mass
following the procedure described in \citet{Glazebrook04} and the initial
mass function given in \citet{Baldry03}. In order to obtain the full
distribution function of the allowed mass, a Monte Carlo simulation
re-sampling the photometric errors is done. We measure  the stellar mass
for this host to be $\log{(M_{\star}/M_\odot)}=11.20\pm0.75$. The large
error bar is due to only having upper limits on the rest frame optical
photometry. 

We find the measured stellar masses for these four hosts to be in
complete agreement with our computed values provided we use the
prescription including the metallicity gradient. If we instead use equation
(1) from \citet{Christensen14}, which assumes a constant offset between
absorption metallicity and emission metallicity, then the agreement is
much poorer with the computed masses in the mean being 1 dex higher than
the measured stellar masses. I.e. our data support the hypothesis that
the stellar masses of GRB-DLA galaxies follow the same prescription as
do QSO-DLA galaxies, and that they have metallicity gradients with a
slope similar to that of QSO-DLAs.

In a related study of a sample of 18 low redshift GRB hosts with
measured emission line metallicities and stellar masses from SED fits,
\citet{Mannucci11} showed that those host galaxies follow the same
M-Z relation as SDSS galaxies, but only after correcting for the high
SFR which is a result of the SFR weighted selection we discussed in
section \ref{histos}.
It therefore appears that the available samples of emission selected
galaxies, GRB selected galaxies, and DLA selected galaxies follow the
same M-Z relations (when corrected for their specific selection
function) and likely are drawn from the same underlying galaxy sample.

In Section \ref{histos} we described how the metallicity offset seen in Figure \ref{vz-dist}
could be understood as a result of selection functions, but from
Figure \ref{zb-corrected} it is seen that the offset could just as well be caused by
the effect of metallicity gradients and different impact parameter
distributions. Our sample covers a range of stellar masses from
$10^{6.7}$ to $10^{11}$ $M_{\odot}$, with a median of $10^{8.5}$
(Table \ref{Tab5}). This median mass is identical to that reported by
\citet{Moller13} for DLA galaxies which, held together with the better
fits using metallicity gradients described above, supports that at
least part of the metallicity offset is a result of different impact
parameter distributions. In that case the shift between the two
samples seen in Figure \ref{zb-corrected} is most easily interpreted as the effect of
different paths through the gravitational potentials.

The interpretation of the observed distribution of data points in
Figure \ref{z-corrected} is therefore complex. Effects of redshift evolution, impact
parameter distributions, metallicity gradients, and differently
weighted selection functions all work to move the data-points, which
causes at least part of the scatter of the relation. 
We here repeat from the conclusions of \citet{Moller13} that in
order to move forward towards an understanding of those objects we
need to identify and understand the sources of the scatter. One of
the sources (redshift evolution) has already been identified. 
\citet{Christensen14} recently found that half of the scatter in their sample was
removed when the effect of metallicity gradients were included. Here
we have proposed that the effect of gravitational well depth could 
be an additional cause of scatter.


\section{Conclusions}
\label{summary}
Most long duration GRB host galaxies display strong intrinsic DLA
absorption systems similar in nature to the intervening 
DLA systems seen in QSO spectra. The GRB host systems are, however, different in two ways: they originate
inside the host galaxies rather than behind them and they are found at much smaller impact parameters. In 
addition they are also reported generally  to have higher HI  column densities and often  to have higher 
metallicities than intervening DLAs at the same redshift.

It is important to establish if those differences simply are a result of two different selection functions 
applied to the same underlying sample of high redshift galaxies, or if the two types of galaxies are truly 
two different populations.

We have here analysed the mass/metallicity/redshift relations of a
complete literature sample of GRB host galaxies and a sample 
of intervening DLA galaxies in order to address this question. We have found that\\
1) The two samples are fully consistent with being drawn from the same
underlying population with a single MZ relation, and a single redshift
evolution of this relation with a break around $z \approx 2.6$. 
GRB hosts are in better agreement with this   
redshift evolution  compared to linear 
evolutions with constant slopes. 
\\
2) There is evidence that the GRB host galaxies have higher
metallicities, but this is most likely a secondary correlation. The
primary correlation is with either impact parameter, with stellar
mass, or, presumably, with a combination of the two. The smaller impact
parameters combined with a metallicity gradient will produce a metallicity
offset, SFR selection bias is predicted to select galaxies of somewhat
larger stellar mass than DLA galaxies which will likewise cause an
offset in metallicity.\\
3) There is weak evidence that the $\Delta v_{90}$-metallicity
relation for the GRB hosts is offset towards larger $\Delta v_{90}$
values, as one would predict since their sightlines pass through a
deeper part of the dark matter halo potential well than a random
sightline to an intervening DLA in a halo of the same mass.

It has been shown previously that QSO-DLAs and Lyman Break Galaxies (LBG) 
are consistent with being drawn from the same underlying population by
two very different selection functions, where QSO-DLAs are drawn from 
the very low-mass end of the LBG population \citep{Moller02}. With the
results presented here we have now added long duration GRB hosts to this
list, which means that we have made another important step towards a global
description of galaxies and galaxy evolution in the early universe.

Since the sample used in this pilot study is limited, it will be quite
feasible to improve the accuracy of all results reported here simply
by increasing the sample size. GRB host are ideally suited to shed
light on the structure of high redshift galaxies. They combine the
data from emission selected galaxies directly with those of absorption
selected galaxies. I.e. we obtain in the ideal case both absorption
metallicity, emission metallicity, stellar mass from SED fits, impact
parameter and $\Delta v_{90}$ for a single galaxy.

\section*{Acknowledgment}
We would like to thank Jason X. Prochaska for providing the HI column
density for the host of GRB 090313 prior to publication. We thank an
anonymous referee for a careful reading of our manuscript and for many
insightful and valuable comments which significantly improved the
presentation of our results.
We thank Karl Glazebrook and Damien Le Borgne for helping with the stellar mass calculation of the host of GRB 090323.
MA thanks Max-Planck-Institut f\"{u}r Astrophysik for hosting her during the initial part of this work. 
JPUF acknowledges support from the ERC-StG grant EGGS-278202. 
This work was funded by an ESO DGDF grant to PM and WF.  
The Dark Cosmology Centre is funded by the Danish National Research Foundation.

\bsp

\label{lastpage}

\end{document}